\documentclass[useAMS,usenatbib]{mn2e}
\usepackage{graphicx}

\title[Transit spectroscopy of GJ 436b with the HST]{Transit infrared spectroscopy of the hot neptune around GJ~436 with the Hubble Space Telescope\footnote{Based on observations with the NASA/ESA Hubble
Space Telescope, obtained at the Space Telescope
Science Institute, which is operated by AURA, Inc.,
under NASA contract NAS 5-26555.}}

\begin{document}
\author[Pont et al. ]{  F. Pont$^{1}$, R. L. Gilliland$^{2}$, H. Knutson$^{3}$, M. Holman$^{3}$, D. Charbonneau$^{3}$\\
$^{1}$ University of Exeter, The QueenÕs Drive, Exeter, Devon, UK EX4 4QJ\\
$^{2}$ Space Telescope Science Institute, 
3700 San Martin Drive,
Baltimore, MD 21218\\
$^{3}$ Harvard-Smithsonian Center for Astrophysics,  60 Garden Street,   Cambridge MA 02138\\
}  
\bibliographystyle{mn2e}



\maketitle

\label{firstpage}

\begin{abstract}

The nearby transiting system GJ 436b offers a unique opportunity to probe the structure and atmosphere of an extra-solar ``hot Neptune''. In this Letter, we present the main results of observations covering two transit events with the NICMOS camera on the Hubble Space Telescope. The data consist of high-cadence time series of grism spectra covering the 1.1-1.9 $\micron$  spectral range. 
We find $R_{pl}=4.04 \pm 0.10\ R_\oplus$ and $R_*= 0.446 \pm 0.011\ R_\odot$ for the planet and star radius, confirming and improving earlier measurements with ground-based photometry and a Spitzer lightcurve at 8 $\micron$, as opposed to a much higher value obtained with the Fine Guidance Sensor on the Hubble Space Telescope. We measure no departure from strict periodicity in the transits to the level of $\sim \!7$ seconds. This strongly disfavours the proposed explanation of the orbital eccentricity of GJ~436b in terms of the perturbation by another close-by planet.
We measure a flat transmission spectrum at the level of a few parts per 10$^4$ in flux, with no significant signal in the 1.4-$\micron $ water band to a level comparable to the maximum amplitude of the effect predicted by planetary atmosphere models. 
 
\end{abstract}



\section{Introduction}

The  planet orbiting around the nearby M-dwarf star GJ 436 is the first known transiting neptune-class planet -- and to date the only one (Butler et al. 2004, Maness et al. 2007, Gillon et al. 2007). The planet has a 2.64 day orbit with  an eccentricity of $0.15$, and a mass of $\sim 23$ Earth masses. Its radius of $\sim 4\ R_\oplus$ implies a primarily icy or rocky composition with an envelope of hydrogen and helium (Adams et al. 2008), a hotter version of Uranus or Neptune. Measurements of the secondary eclipse in the mid-infrared with the Spitzer Space Telescope indicate a photospheric temperature in the vicinity of 700 K (Deming et al. 2007).  

A second planet has been hypothetized to explain the high orbital eccentricity (e.g. Ribas et al. 2008a). This planet would induce variations in the transit timing detectable with sufficiently precise transit photometry.

Spectroscopic time series during transit can also provide information on the transmission spectrum of the planetary atmosphere: in wavelengths where its atmosphere is more opaque, the planet appears larger across the stellar disc and the transit is slightly deeper. This effect was first detected for the transiting system HD 209458 by Charbonneau et al. (2002), and recently used to obtain an extensive characterisation of the atmospheric transmission spectrum of HD 189733b (Pont et al. 2007, Swain et al. 2008).  

We here present the analysis of two sequences of spectra obtained with the NICMOS camera on the Hubble Space Telescope (HST) with the G141 grism, during two transits of the GJ 436 system on November 11 and December 15, 2007. These data pertain to the issues mentioned above: the resulting precise time series in total flux provide very accurate constraints on the stellar and planetary radius, and determine the timing of the two transits to a few seconds, the spectroscopic time series constrains the planetary atmosphere transmission spectrum.

\section{Data and Reduction}

The images were taken with the G141 grism on the NICMOS camera, sampling the 1.1-1.9 $\micron$ spectral interval. The first HST visit consists of 937 images taken between JD=2454415.48 and JD=2454415.71, the second visit of 917 images between 2454449.85 and 2454450.08. The images are 1.993-second
exposures obtained on 11-second centers.  Since GJ 436 is not situated in the continuous viewing zone of HST, data were acquired during only half of the orbits. The timing of the observations is such that  orbit 3 (of 4) of each visit occurs during the transit, sampling the transit centre and egress. Orbit 2 and 4 are used to set the out-of-transit baseline and calibrate the instrumental effects. As in previous similar observations, orbit 1, during which the telescope and instrument settle to the new mode, cannot be used at this level of accuracy.
 We used the same defocus setting as in previous NICMOS grism observations of bright transiting planets. The highest signal peak on the detector is 87$\,$000 electrons, which corresponds to 43\% of the 215$\,$000 full well-depth for the NIC3 detector.

\subsection{Flatfielding and flux extraction}

The standard pipeline was used to perform bias level, zero read subtraction, non-linearity correction, dark subtraction and cosmic ray removal.  Wavelength-dependent flat-field corrections were computed by interpolating between a set of calibration flatfields taken in different filters.

To calculate the flux, we define an extraction box out to a specified intensity level chosen by running several trials and selecting the one that returns the smallest time series rms over out-of-transit orbits 2 and 4 of each visit. We interpolate spatially over a small set of dead and extra noisy pixels. A global sky level is then evaluated and subtracted for each image. 

\subsection{External parameter decorrelation}

Even though the instrumental setup, pointing and observing conditions were kept very stable, the photon signal-to-noise ratio is so high in our data set that it is important to correct even tiny instrumental effects. 

As in Pont et al. (2007), we removed these effects to first order with multi-linear decorrelation.
We track four external variables for each exposure: the drift of the spectrum in $x$ and $y$ coordinates on the detector, its rotation and width change. The shift in $x$-position was evaluated by cross-correlating extracted 1-d spectra. The changes in $y$-position, width and rotation were defined by fitting a set of 1-d gaussians in the cross-dispersion direction along the first-order spectrum.  The amplitude of the changes was $\pm 0.04$ pixels in position, $\pm 0.03$ pixels in width, and $\pm 0.05$ degrees in rotation in the first visit, and about twice as large in the second visit. To these four vectors we add a linear time vector as a fifth decorrelation parameter (to account for a constant drift over the whole observing sequence). 
For each of the two visits, we calibrate the corrections from a multi-linear regression of the flux residuals against the value of the external parameters for the second and fourth orbit (the first orbit is discarded and the third consists of in-transit data).

This external parameter decorrelation decreases the r.m.s. of the intensity in 1-minute time intervals from $3.27 \cdot 10^{-4}$ to $1.97 \cdot 10^{-4}$ for the first visit and $5.68 \cdot 10^{-4}$ to $2.07 \cdot 10^{-4}$ for the second. Contrarily to the case of ACS data  on HD~209458 (Pont et al. 2007), the latter values are still 2-3 times larger than the photon-noise limit ($0.74 \cdot 10^{-4}$). 


Nevertheless, the decorrelation produces satisfactory results for the data of the first visit, as shown a posteriori by the residuals around a transit lightcurve fit. The results are less well-behaved for the second visit. The reason is that several external parameters take values during the third orbit (the in-transit orbit), that are not covered during the second and fourth orbits used for decorrelation. Therefore, the decorrelation is performed by extrapolating the dependence between parameters and residuals outside the range acually sampled. 


\subsection{Correlated noise analysis}

\label{noise}

Noise on transit time series has a very different impact depending on its covariance properties, as discussed in Pont, Zucker \& Queloz (2006). Noise correlated in time and frequency, such as that produced by telescope and detector systematics, produces much higher uncertainties on the final results than uncorrelated noise sources such as the photon shot noise, because  correlated errors do not average out quickly with higher number of exposures. 

We model the noise as a combination of purely random and entirely correlated components, following the approach of Pont et al. (2006): 
\[ \sigma^2_{\rm tot}=\sigma^2_{\rm w}+\sigma^2_{\rm t}+\sigma^2_{\nu} \]
with $\sigma_{\rm w}$ the purely white noise component, $\sigma_{\rm t}$ and $\sigma_{\nu}$ the components correlated in time and frequency respectively.  The coefficients are estimated by ensuring $\sigma^2= (\sigma^2_w+\sigma^2_\nu)/N+\sigma^2_t$ in 15-minute time bins, and  $\sigma^2= (\sigma^2_w+\sigma^2_t)/N+\sigma^2_\nu$ in 14-pixel bands along the dipsersion direction (14 pixel is the approximate PSF size). We find $\sigma_{\rm tot}=4.7 \cdot 10^{-4}$, the total r.m.s. of individual intensity values, breaks down into:
$\sigma_{\rm w}=3.9 \cdot 10^{-4}, \sigma_{\rm t}=2.3 \cdot 10^{-4}, \sigma_{\nu}=1.0 \cdot 10^{-4}$. The white-noise term averages out when integrated over large number of exposures. For global parameters like the planet-to-star radius ratio, the second terms (noise with time corelation) will dominate the error budget. We thus expect the effect of correlated noise to amount to $\sim 3 \cdot 10^{-4}$ on the transit depth, corresponding to a precision of 0.6\% on the radius ratio of the system.
For the transmission spectrum, the most relevant source of noise is the last one (noise with wavelength correlation). The noise budget is $\sim \! 1.5 \cdot 10^{-4}$ on transmission spectral features including white and correlated noise. 

\section{Integrated-light lightcurve analysis}

\subsection{Transit parameters}

\begin{figure}
\resizebox{8cm}{!}{\includegraphics{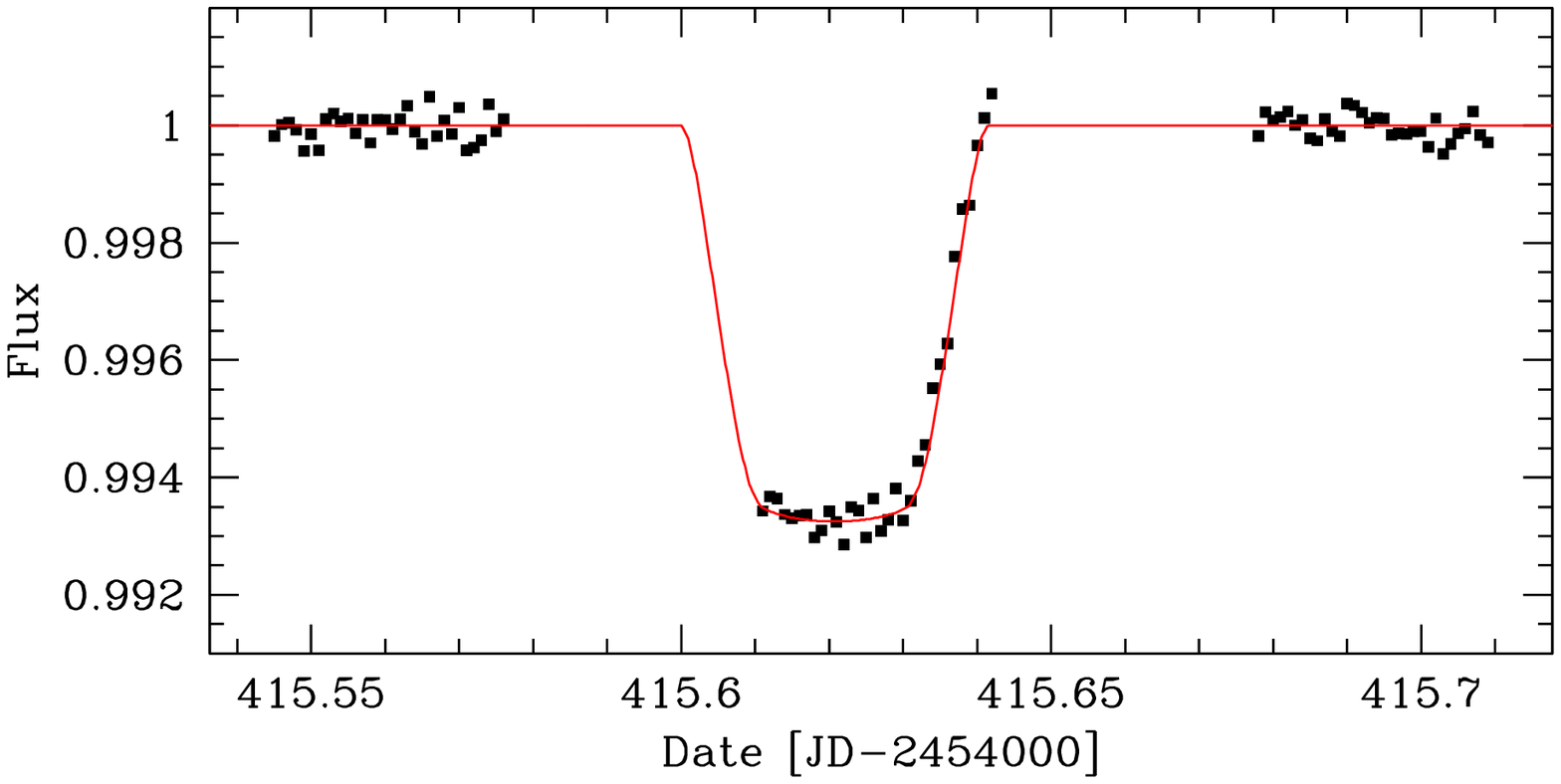}}
\resizebox{8cm}{!}{\includegraphics{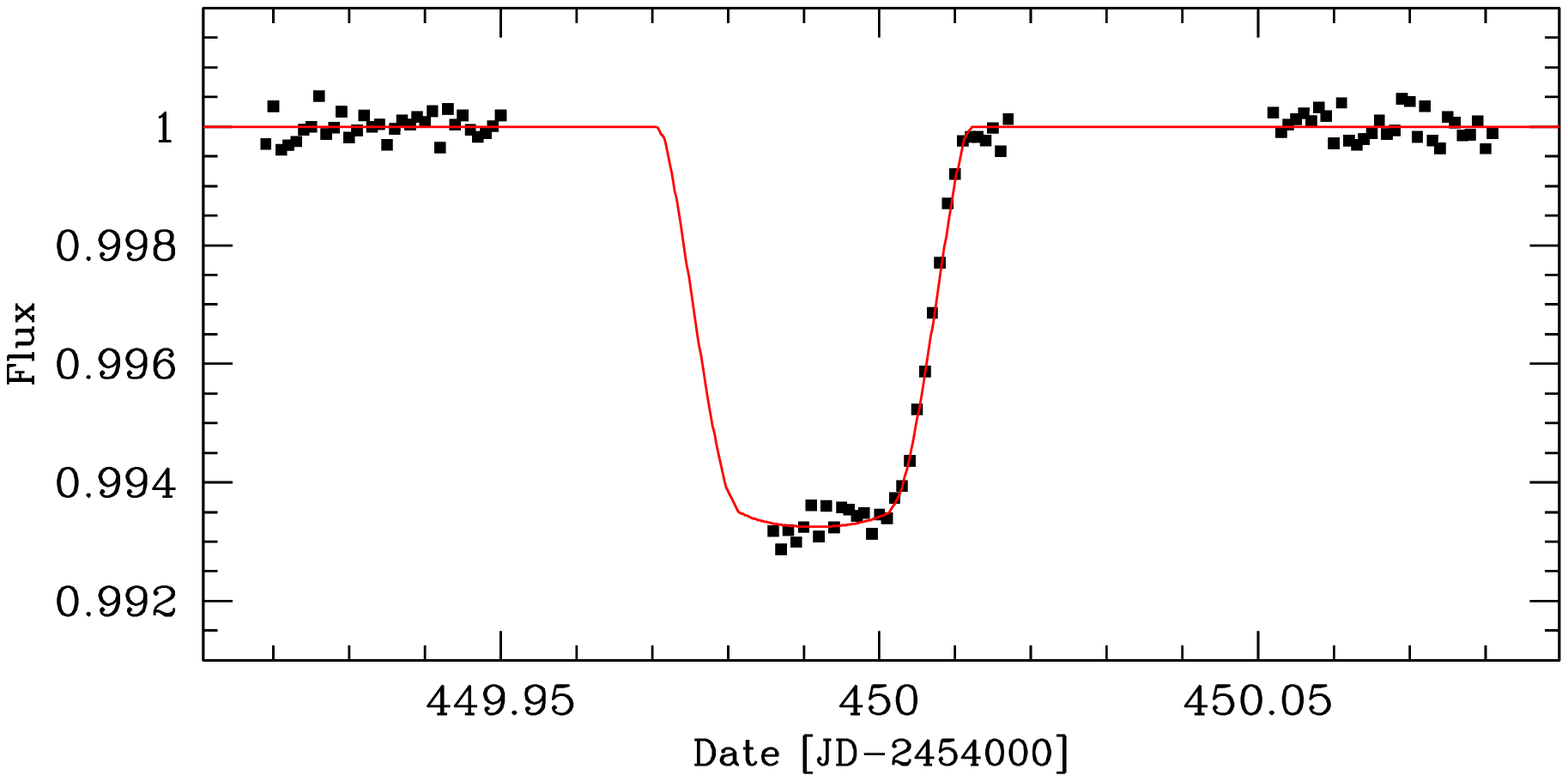}}
\caption{NICMOS integrated-light time series (decorrelated for instrumental effects), in  1-minute bins (5-6 data points per bin), with best-fit transit model light curve, for the two HST visits. The rms around the curve is $1.97\cdot 10^{-4}$ for the first visit and $2.07\cdot 10^{-4}$ for the second}
\label{lc}
\end{figure}

The lightcurve integrated over the whole spectrum is plotted in Fig.~\ref{lc}. The residuals have the same amplitude and correlation properties inside and outside the transit, indicating no significant deviations from the signal expected for the transit of an opaque body in front of an M-dwarf star. In particular, no indication of the transit of another, smaller body in front of the star is seen, nor in-transit residuals that could indicate that the planet is occulting a large star spot (as was the case for HD 189733 in Pont et al. 2007). 

Since our data set covers  the transit egress twice and never the ingress, it is not suited to measure the position of the mean epoch of the transit. We therefore adopt the epoch and period of the transit signal obtained on five partial transits with the Fine Guidance Sensor of the HST by Bean et al. (2008; B08): $P=2.643904$ days, $T=2454455.27924$ HJD.  We also adopt the eccentricity and argument of periastron from the radial-velocity orbit. At the level of accuracy reached by our data, it is important to use a precise limb-darkening profile, even though  the effect of limb darkening is somewhat lower in the near infrared than at shorter wavelengths. 
To calculate the limb-darkening parameters, we used a model selected from a grid of Kurucz atmosphere models with an effective temperature of 3500 K, log(g) = 5.0, a turbulent velocity of 1 km/s, and [Fe/H] = 0.0. We calculate limb darkening numerically on 17 points along a chord crossing the stellar disc, and fit the four-coefficient non-linear law of Claret (2000). We calculate the limb-darkening coefficients for the central wavelengths of 114 one-pixel intervals along the spectral dispersion of the NICMOS spectra, then compute the intensity-weighted mean of the coefficients for the integrated-light lightcurve. We find  (1.533, $-$2.234, 1.913, $-$0.643) for the four coefficients.

We fit a transit curve model using the Mandel \& Agol (2002) algorithm on the NICMOS lightcurve, obtaining $R_{pl}/R_*=0.0831 \pm 0.0005$, $T_{tr}=0.0317 \pm 0.0004$ days and $b=0.850 \pm 0.007$ for the geometric transit parameters (radius ratio, transit duration\footnote{transit time of the center of the planet across the stellar disc} and impact parameter). The uncertainties are calculated using the method of Pont et al. (2006), accounting for the presence of correlated noise in the data. Fitting the two transits separately yields $R_{pl}/R_*=0.0832 \pm 0.0006$, $T_{tr}=0.0317 \pm 0.0004$  and $b=0.857 \pm 0.013$ for the first visit and  $R_{pl}/R_*=0.0830 \pm 0.0006$, $T_{tr}=0.0320 \pm 0.0004$  and $b=0.843 \pm 0.014$ for the second visit. The two sets of values are compatible within their error bars.

We can then iterate on the transit epoch and period to attain a self-consistant solution. With the geometric transit parameters fixed, we free only the two transit timings for the NICMOS visits, finding: $T_1=2454415.62074 \pm 0.00008$ and $T_2=2454449.99141 \pm 0.00008$. These values are not significantly different from those obtained by extrapolating from the epoch and period found by B08 ($5\pm 7$ seconds earlier and $2 \pm 7$ seconds later) and used in the first fit. Further iterating is thus unnecessary.

\subsection{Transit timing}

\begin{figure}
\resizebox{8cm}{!}{\includegraphics{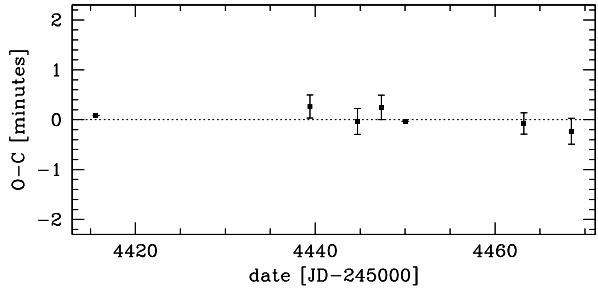}}
\caption{Transit timings for our two transits (smaller error bars) and the five transits with FGS from Bean et al. (2008). The dotted line shows the expectation for a strictly periodic signal compatible with the earlier Spitzer transit. }
\label{tt}
\end{figure}

Figure~\ref{tt} shows the transit central times of the NICMOS and FGS data, as residuals compared to the solution of B08, that also takes into account the earlier Spitzer and ground-based measurements. The transit signal is found to be precisely periodic to the level of a few seconds. Such stability severely constrains the presence of a second massive planet with a period comparable to that of GJ 436b, as invoked for instance by Ribas et al. (2008a) to account for radial velocity residuals and the orbital eccentricity.  This scenario (and subsequent variations to account for new observations, e.g. Ribas et al. 2008b) predicts oscillations of several minutes in the transit times, which, barring highly unlikely coincidences, are excluded by the NICMOS data.

We performed numerical integrations similar to those in Agol et al. (2005) and Holman et al. (2005) to estimate the amplitude of transit timing variations that would be induced by an additional planet in the GJ~436 system.  A perturbing planet with mass 0.01 $M_\oplus$ in either the 2:1 or 3:1 mean motion resonance will generate r.m.s. transit timing variations of $\sim$7 seconds, assuming the perturbing planet shares the orbital plane, argument of pericenter, and eccentricity (e=0.15) of GJ~436b.    Likewise, a ~5 $M_\oplus$ planet in a similar but non-resonant orbit between the 2:1 and 3:1 mean motion resonance will generate $\sim$7 seconds r.m.s. timing variations. 
 
Bean \& Seifahrt (2008) discuss the possible location in the $M_p - a$ plane of a second planet accounting for the eccentricity of GJ~436b (see their figure~1). Our data, while not strictly excluding all scenarios, further narrows the parameter space available. If GJ 436b's unusual eccentricity is not the result of perturbations from a second planet in the system, we must consider the possibility that the planet's eccentricity is primordial and we have simply underestimated the circularization time-scale for this system.  This would require an unusually high value for the planet's tidal Q factor, but this factor is generally poorly constrained for extrasolar planets (Ogilvie \& Lin 2004, Wu 2005).


\begin{table}
\centering
\begin{tabular}{l r}
\hline
Period [days] & 2.643904 (from B08) \\
Radius ratio & $0.0831 \pm 0.0005\ \, $ \\
Impact parameter & $0.850 \pm 0.007\ \ \ $ \\
Transit duration [days] & $0.0317 \pm 0.0004\ \ \ \ \,  $  \\
Transit timing [HJD] & $2454415.62074 \pm 0.00008$ \\
                             & $2454449.99141 \pm 0.00008$ \\
Star radius [R$_\odot$] & $0.446 \pm 0.011\ \ \ $\\
Planet radius [R$_\oplus$] & $4.04 \pm 0.10\ \ \  \ \,$ \\ \hline

\end{tabular}
\caption{New and updated parameters of the GJ 436 system}
\end{table}

\subsection{Planetary radius}

Adopting $M_*=0.452 \pm 0.013 $ from Torres et al. (2007), the transit parameters found from the NICMOS data imply $R_*=0.446 \pm 0.011\ R_\odot$ for the stellar radius and $R_{pl}=4.04 \pm 0.10\ R_\oplus\ (25750 \pm 650$ km) for the planetary radius. The systematic uncertainties due to stellar evolution models (for $M_*$) are not included in the error estimates.


Our value for the planetary radius is in agreement with initial ground-based estimates and the value derived from the Spitzer coverage at 8 $\micron$ ($R_{pl}\simeq 4.0 -4.3\ R_\oplus $ ). We do not confirm the significantly higher value obtained by B08 ($R_{pl}=4.90^{+0.45}_{-0.33}$ $R_\oplus$) from the FGS lightcurves. B08 discuss several possible explanations for either the Spitzer or the FGS radius to be off by $\sim 2$ sigmas. In particular, the FGS lightcurve is built from five different partial transits, with relative shifts between the different transits as free parameters.  Unrecognized time-dependent effects (instrumental systematics as well as stellar variability and spots) could thus modify the shape of the transit. The effect of stellar limb darkening is also much stronger in the visible wavelengths detected by FGS than in the near infrared, and more difficult to calculate from models -- especially for an M-type spectrum. The effect of stellar variability and spots is also larger for shorter wavelengths. Given these factors and the higher total signal-to-noise ratio of the NICMOS data, the radius value that we find suggests that the FGS radius is indeed overestimated. 


\section{Spectroscopic analysis}

The primary objective of our observations was to provide broad constraints on the transmission spectrum of the planetary atmosphere in the $1.1-1.9\ \micron $ range, in particular around the water bands near 1.4 $\micron$.  


Models for this planet assuming a primarily hydrogen-dominated atmosphere with solar metallicity predict that water will exist in gas phase in the upper atmosphere, as this region is too hot for water vapour clouds and too cool for water to be thermally dissociated.  At the predicted abundances this water vapour produces a strong absorption signal at near-IR wavelengths, including one band at 1.4 $\micron$ that falls in the center of our spectrum.

We can estimate in an approximate way the size of the expected absorption signal by calculating the atmospheric scale height, which is given by:
$
H= k T/(g \mu)
$
where $T$ is the temperature, $g$ is the surface gravity, and $\mu$ is the mean molecular weight of the atmosphere.  For GJ~436b, assuming a temperature of 710~K (Deming et al. 2007, Demory et al. 2007), a surface gravity of 1280 cm$\,$s$^{-1}$ (Butler et al. 2004, Torres 2007), and an atmosphere of molecular hydrogen, this would correspond to a scale height of 230~km.  The change in the depth of the transit is proportional to the additional area occulted, typically approximated as 10$\cdot$H (Seager \& Sasselov 2000), so the total size of the signal would be 2 x 10-4.  This is consistent with predictions from full 1D radiative transfer atmosphere models for GJ 436b (E. Miller-Ricci \& S. Sasselov, private communication), which indicate that the predicted signal from water absorption in our selected bandpass should be  0.6-1.1 $\times 10^{-4}$, where the lower end of this range corresponds to an atmosphere with 30$\times$ solar metallicity and the upper end is for an atmosphere with solar metallicity (see Miller-Ricci et al. 2008 for a full description of the methodology used for these models).  Uniformly increasing the fraction of heavy elements in this planet's atmosphere does not significantly alter the relative depths of the various absorption features, but it does produce a net decrease in the strength of those features as the increased value of $\mu$ reduces the atmospheric scale height. 

To recover the wavelength-dependent information from the NICMOS data, we repeated the extraction along columns on the CCD perpendicular to the direction of dispersion. We built a differential indicator of the intensity of absorption in the 1.4-$\micron $ water band, by computing the difference between the flux in a 1.34-1.53 $\micron$ passband with the mean of the flux in two side bands, 1.21-1.30 and  1.55-1.64 $\micron$. We correct for the different limb-darkening coefficients in these passbands with the Kurucz models.  


The presence of a planetary atmosphere with excess opacity in the water band would leave a transit-like signature in the run of this indicator with time, because the planet would appear larger within the wavelength of the band than outside.    We find a flux {\it excess}  of $+1.42 \pm 0.89 \times 10^{-4}$ in the water passband during the transit. The uncertainty is estimated from the variance of the time series of the water indicator, and does not include the uncertainties in the corrections of instrumental systematics. Based on the noise analysis of section~\ref{noise}, we estimate these at $\sim \!1.0 \times 10^{-4}$.  

To sum up, we find a $\sim \!1$-sigma signal in the direction opposite to the expected water absorption band, with a comparable amplitude. 
Thus, although  spectral signatures much larger than that predicted by the models are ruled out, the NICMOS data is not sufficient to measure spectral features of the planetary atmosphere.



\begin{figure}
\resizebox{8cm}{!}{\includegraphics{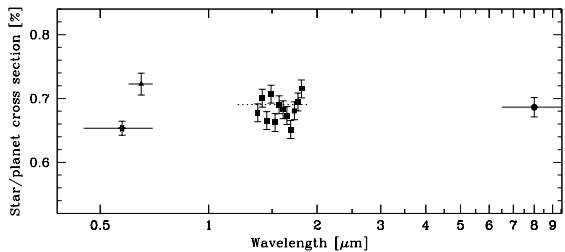}}
\caption{Planet cross-section (in percentage of the stellar disc covered) as a function of wavelength for GJ~436b. The NICMOS data (squares) is binned over 5-pixel intervals across the spectral dispersion direction. The short-wavelength points come from a collection of ground-based data (Shporer et al. 2008, triangle) and the B08 data (re-fitted with our orbital parameters). The long-wavelength point is the 8.0-$\mu$ m Spitzer transit (Southworth 2008). Error bars for space data do not include the uncertainties on the correction of instrumental systematics. }
\label{spec}
\end{figure}




We also built a low-resolution transit spectrum from our data, to check whether any feature in the planet transmission spectrum was visible above the uncertainties introduced by residual instrumental systematics.  We binned the column-by-column wavelength extraction in 5-pixel bands (0.04-$\micron$ passbands), and fitted a transit shape with all parameters fixed to the best-fit values and theoretical limb-darkening coefficients, except the transit depth. Wavelength-dependent systematics in the time series were corrected for each passband to first order with a five-parameter description: two quadratic functions of HST orbital phase before and after the moment when the telescope enters the Earth's shadow, and a zero-point drift. We retain only the data when the covariance between the decorrelation parameters and the transit depth was small, restricting the results to the $1.35\ \micron \!<\! \lambda \!<\! 1.85\ \micron $ interval and the first HST visit. Outside this wavelength interval, the resulting spectrum depends significantly on the parameters used in the decorrelation. For the second visit, the wavelength-dependent decorrelation is not stable, because the pointing and focus of the telescope took values during the in-transit orbit that are incompletely sampled by the orbits before and after the transit.

Figure~\ref{spec} plots the results in terms of transit cross-section ($R_{pl}^2/R_*^2$), together with results from ground-based observations (Shporer et al. 2008), the Spitzer 8.0$-\micron$\ measurement (Southworth et al. 2008), and the B08 result re-fitted with our transit parameters. The dotted line shows the best-fit cross-section from the integrated-light NICMOS data. The NICMOS spectrum has a rms of $1.9\cdot 10^{-4}$, compared to an expected noise of $1.4\cdot 10^{-4}$ (including $1.0 \cdot 10^{-4}$ noise correlated in wavelength). 
The data can be explained by a constant radius over all the wavelength range, and there is no compelling evidence at this point for spectral features that could be attributed to the effect of the planetary atmosphere at the level of a few parts per $10^4$.

\section{Conclusion}

Two transits of the hot neptune GJ 436b were measured in very high signal-to-noise ratio grism spectroscopy in the 1.1$-$1.9 $\micron$  range with NICMOS on the HST. The transit shape indicates a size of $R_{pl}=4.04 \pm 0.10\ R_\oplus $ for the planet (assuming $M_*=0.452 \pm 0.013$ $M_\odot$), confirming the values obtained with the Spitzer 8-$\micron$  data, rather than the higher radius from FGS/HST data in the visible. The higher FGS value is probably due to instrumental systematics, unexpected values of limb darkening, or irregularities on the stellar surface.

No significant departures from a pure, strictly periodic transit signal are present beyond the level of time-correlated residuals from instrumental systematics ($\sim 10^{-4}$ in flux), indicating the absence of strong intensity fluctuations on the surface of the star crossed by the planet, and of transit timing variations on the scale of a few dozen seconds. This is a strong argument against a second close-in planet producing the perturbations resonsible for the eccentricity of the orbit of GJ 436b. 

We analyse the wavelength dependence of the transit depth in search of features introduced by the planetary atmosphere. We find no such feature at the level of a few parts per $10^4$, and measure a flux excess of $+1.42 \pm 0.89 \cdot 10^{-4}$ in the water vapour absorption  band around 1.4-$\micron$ (where models predict a signal of $10^{-4}$ or lower in the opposite direction). 
Therefore, the data excludes an anomalously large signal in transmission spectroscopy in this band, but is not sufficient to measure the actual size of the water absorption feature. 
Our ability to remove systematics from the data is ultimately the limiting factor in this analysis; if we are to achieve a detection of water vapour in this planet's atmosphere we will have to improve our corrections for these effects in future analyses of HST observations and reach a noise level much closer to that of the photon-noise-limited precision for this system.

\section*{Acknowledgements}

{\it The authors wish to thank Eliza Miller-Ricci and Sara Seager for atmosphere models provided in advance of publication and insightful comments,  Michel Mayor for support with the observing proposal,  Daniel Fabrycky for helpful discussions about transit timing variations in the GJ~436 system, and the anonymous referee for numerous helpful comments.}

\section*{Bibliography}

Adams E.R., Seager S., Elkins-Tanton L., 2008, ApJ 673 1160\\
Agol E., Steffen J., Sari R., Clarkson W., 2005, MNRAS 359, 567\\
Bean  J. L., Benedict  G. F., Charbonneau  D. et al.  2008, A\&A 486, 1039\\
Bean  J. L., Seifahrt, A., 2008, A\&A Letters 487, L25\\
Butler R. P., Vogt S. S., Marcy G. W. et al., 2004, ApJ 617, 580\\ 
Charbonneau D., Brown T. M. Noyes R. W., Gilliland R. L., 2002, ApJ 568, 377\\
Claret A., 2000, A\&A 363, 1081\\
Deming D., Harrington J., Laughlin G., Seager S., Navarro S. B., Bowman W. C., Horning K. , 2007, ApJ 667, 199\\
Demory B.-O., Gillon M., Barman T., 2007, A\&A 475, 1125\\
Gillon M., Pont F., Demory B.-O. et al., 2007, A\&A Letters 472, L13\\
Holman M. J., Murray N. W., 2005, Science 307, 1288
Mandel K., Agol E., 2002, ApJ 580, 171\\
Maness  H. L., Marcy  G. W., Ford  E. B. et al., 2007, PASP 119, 90\\
Miller-Ricci E., Seager S., Sasselov D., 2008, ApJ, in press, arXiv:0808.1902\\
Ogilvie G.~I.,  Lin D.~N.~C., 2004, ApJ 610, 477\\
Pont F., Zucker S., Queloz D., 2006, MNRAS 373, 231\\
Pont F., Gilliland R. L., Moutou C. et al., 2007, MNRAS 385, 109\\
Ribas I., Font-Ribera A., Beaulieu J.-P., 2008a, ApJ 677, 59\\
Ribas I., Font-Ribera A, Beaulieu J.-P. et al., 2008b, in IAU Symp. 253, eds. F. Pont et al., in press\\ 
Seager S., Sasselov D.D., 2000, ApJ 537, 9\\
Shporer, A., Mazeh, T., Winn, J.~N., Holman, M.~J., Latham, D.~W., Pont, F., Esquerdo, G.~A., 2008, A\& A, arXiv:0805.3915\\
Southworth, J., 	2008, MNRAS 386, 1644\\
Swain M., Vasisht G., Tinetti G., 2008, Nature 452, 329 \\
Tinetti G., Vidal-Madjar A., Liang M. et al., 2007, Nature 448, 169 \\
Torres, G., 2007, ApJ 671, 65\\
Wu Y., 2005, ApJ 635, 674
\end{document}